\documentclass[useAMS,usenatbib]{mn2e}

\usepackage{epsfig}
\usepackage{threeparttable}
\usepackage{amsmath}
\usepackage{amssymb}
\usepackage[british]{babel}
\usepackage{times}

\newcommand{\Ha}{$H_{\alpha}$}
\newcommand{\Hb}{$H_{\beta}$}
\newcommand{\Hg}{$H_{\gamma}$}
\newcommand{\Hd}{$H_{\delta}$}
\newcommand{\pg}{PG\,1018--047}
\newcommand{\ion}[2]{{\rm{#1}}\,{\sc{#2}}}

\title[PG\,1018--047: the longest period subdwarf B binary]{PG\,1018--047: the longest period subdwarf B binary}

\author[J. Deca et al.]{J. Deca$^{1,}$\thanks{E-mail: jan.deca@wis.kuleuven.be}, T. R.
Marsh$^{2}$, R. H. \O{stensen}$^{3}$, L. Morales-Rueda$^{4}$, C. M. Copperwheat$^{2}$,
\newauthor  R. A. Wade$^{5}$, M. A. Stark$^{6}$, P. F. L. Maxted$^{7}$, G. Nelemans$^{8}$ and U. Heber$^{9}$\\
$^{1}$Centrum voor Plasma-Astrofysica, Katholieke Universiteit Leuven, Celestijnenlaan 200B, B-3001, Leuven, Belgium\\
$^{2}$Department of Physics, University of Warwick, Coventry CV4 7AL, UK\\
$^{3}$Instituut voor Sterrenkunde, Katholieke Universiteit Leuven, Celestijnenlaan 200D, B-3001 Leuven, Belgium\\
$^{4}$Symetrica Security Ltd. Phi House, University of Southampton, Science Park, SO16 7NS, UK\\
$^{5}$Department of Astronomy and Astrophysics, Pennsylvania State University, University Park, PA 16802, USA\\
$^{6}$Department of Computer Science, Engineering \& Physics, University of Michigan-Flint, Flint, MI 48502, USA\\
$^{7}$Astrophysics Group, Keele University, Keele, Staffordshire ST5 5BG, UK\\
$^{8}$Department of Astrophysics, IMAPP, Radboud University Nijmegen, P.O. Box 9010, 6500 GL, Nijmegen, The Netherlands\\
$^{9}$Dr. Karl Remeis-Observatory \& ECAP, Astronomical Institute, Friedrich-alexander University Erlangen-Nuremberg, Sternwartstr. 7, 96049 Bamberg, Germany}
\begin{document}

\date{Accepted 0000 month 00. Received 000 month 00; in original form 0000 month 00}

\pagerange{\pageref{firstpage}--\pageref{lastpage}} \pubyear{2011}

\maketitle

\label{firstpage}

\begin{abstract}
About 50\% of all known hot subdwarf B stars (sdBs) reside in close (short period) binaries, for which common envelope ejection is the most likely formation mechanism. However, \citet{han2003} predict that the majority of sdBs should form through stable mass transfer leading to long period binaries. Determining orbital periods for these systems is challenging and while the orbital periods of $\sim$100 short period systems have been measured, there are no periods measured above 30 days. As part of a large program to characterise the orbital periods of subdwarf B binaries and their formation history, we have found that \pg\ has an orbital period of 760 $\pm$ 6 days, easily making it the longest period ever detected for a subdwarf B binary. Exploiting the Balmer lines of the subdwarf primary and the narrow absorption lines of the companion present in the spectra, we derive the radial velocity amplitudes of both stars, and estimate the mass ratio $M_{\rm MS}/M_{\rm sdB} = 1.6 \pm 0.2$. From the combination of visual and infrared photometry, the spectral type of the companion star is determined to be mid K. 
\end{abstract}


\section{Introduction}


Subdwarf B stars (sdBs) are core-helium burning stars with thin hydrogen envelopes. They are situated between the main sequence and the white dwarf cooling track at the blueward extension of the horizontal branch, the so-called Extreme or Extended Horizontal Branch \citep{ heb1984, heb1986,saf1994}. Subdwarf B stars have colours and spectral characteristics corresponding to those of a B star, but the Balmer lines are abnormally broad for the colour compared with population I main-sequence B stars due to their high surface gravities ($\log g \simeq 5.0 - 6.0$). Subdwarf B stars have a typical mass of 0.5$\mathrm{\,M_{\odot}}$ \citep{heb1984} and can be found in all Galactic populations. They are thought to be the dominant source for the UV-upturn in early type galaxies \citep{fer1991, bro2000}.  A fraction of sdBs pulsate \citep{cha1996, kil1997}, giving great opportunities to derive fundamental parameters (e.g. the stellar mass) and study their internal structure in detail \citep{gre2003, fon2008, ost2009, ost2010}. Subdwarf B stars are also suggested to be very useful as age indicators using evolutionary population synthesis \citep{bro1997}, or as distance indicators \citep{kil1999}. Since a large fraction of sdBs are members of binary systems \citep{max2001} and because they are intrinsically bright and ubiquitous, they are therefore an ideal population in which to study binary star evolution. For a comprehensive review on hot subdwarf stars we refer the reader to \citet{heb2009}.


\citet{han2003} describe in detail the formation and evolution of sdBs by using binary population synthesis models. They find that sdBs form via five main evolutionary channels: the first and second common envelope channels, the first and second stable Roche lobe overflow channels and the helium white dwarf merger channel. This last channel is the only one that results in the formation of single sdBs. They find that the contribution of the second Roche lobe overflow channel is not significant, leaving only three channels to form sdB binaries. Each of these three binary formation channels predicts a different orbital period distribution for the population of sdBs. The binaries formed through the first common envelope channel should display orbital periods between 0.5 and $\sim$40 days and the companions to the sdBs will be main sequence stars. Binaries formed via the second common envelope channel are expected to have white dwarf companions and their range of orbital periods will be wider, extending further into the short periods but not to long periods. Note as well that these common envelope phases are not very well understood. \citet{nel2000} have concluded from the observed double white dwarf population that its outcome may not always be a strong reduction of the orbital separation.
Finally, sdB binaries formed through the first stable Roche lobe overflow channel will have main sequence companions and will display orbital periods between 0.5 and $\sim$2000 days. 

\citet{han2003} conclude that their set 2 of simulations is the model that best describes the observed sample of short period sdB binaries \citep{mor2003}. In this particular model (and also in 9 out of the 12 models they describe) the majority of sdB binaries, between 60 and 70 percent of the total, are formed via the first stable Roche lobe overflow channel. At the same time, this is the channel most affected by observational selection effects decreasing the number of observable sdB binaries formed through this channel. These observational effects are primarily that sdBs with companions that are brighter than the sdB itself will not be identified as sdBs at all. The second effect has to do with observational limitations: it is easier to detect radial velocity variations from a short orbital period system than from a long one, as these are smaller and take longer to determine in the second case.

Despite extensive observational work, not a single system has been found in this long-period regime (first stable RLOF channel), whereas at present $\sim$100 sdB binaries with short periods are confirmed (\citealt{gei2011} and \citealt{cop2011}). The orbital periods are mostly below 1 day, with a median period of 0.61 days.  
However, in this work we will report on \pg, the first truly long-period subdwarf B binary. While this may be the product of binary evolution as suggested by the \citet{han2002, han2003} models, we will find that it could also be the remnant of a hierarchical triple, as recently outlined by \citet{CW2011}. 


Our target, PG\,1018--047 has an apparent visual (Str\"omgren) magnitude $m_{y} = 13.32$ and was discovered as an ultraviolet-excess stellar object in the Palomar-Green Survey \citep{GSL1986}. It was subsequently observed by \citet{max2001} to check for radial velocity variations. Although weak spectral features from a late-type companion are visible in the spectrum, \citet{max2001} did not find any significant radial velocity shifts using their variability criteria. This led to the conclusion that PG\,1018--047 is probably not a binary with a short orbital period. The presence of the companion in the spectrum prompted continued follow-up of the system in order to determine how such a binary could have formed in the first place.  
We present the results after more than a decade of monitoring.\\

\section{Observations and reduction}

We have observed PG\,1018--047 spectroscopically with several different instrument setups over a period of ten years. In Table \ref{tbl:obs} we summarize the observing dates, the setup used in each case, the wavelength range covered and the number of spectra obtained during each epoch. The data were obtained using the Isaac Newton Telescope (INT), William Herschel Telescope (WHT)\footnote{Both the INT and WHT belong to the Isaac Newton Group of Telescopes (ING).} and Nordic Optical Telescope (NOT) on the Island of La Palma, the Radcliffe telescope at the South African Astronomical Observatory (SAAO) and the Hobby-Eberly Telescope (HET) located at the McDonald Observatory in Texas. The different instrument setups were as follows:


For the \textbf{INT-R}ed spectra the intermediate dispersion spectrograph (IDS) was used. It is a long-slit spectrograph mounted on the Cassegrain Focal Station of the INT. The 500\,mm camera together with the high resolution R1200R grating and a windowed Tek5 CCD centered in $\lambda=6560\,\mathrm{\AA}$ covered the $H_{\alpha}$ region. A 1\,arcsec slit was used.

\textbf{INT-B}lue: The INT with the IDS, equipped with the 235 camera, the R1200B grating and a windowed EEV10 CCD was used to obtain these blue spectra. The 2002-2007, 2008-2009 spectra were centered on respectively $\lambda=4348\,\mathrm{\AA}$ and $\lambda=4505\,\mathrm{\AA}$, covering as many Balmer lines to the blue as possible, including $H_{\beta}$ at 4861.327\,\AA. For all exposures a 1\,arcsec slit was used.

The \textbf{SAAO} data were obtained using the Radcliffe 1.9\,m telescope together with the grating spectrograph plus the SITe back-illuminated CCD. Grating 4, with 1200 grooves per millimeter was used to obtain spectra covering $H_{\gamma}$ and $H_{\beta}$ with a dispersion of 0.5\,\AA/pix and a resolution of 1\,$\mathrm{\AA}$ at 4600\,\AA. The slit width varied from 1.2 to 1.5\,arcsec depending on the seeing.

\textbf{WHT-R}ed: The WHT was equipped with the double arm Intermediate dispersion Spectrograph and Imaging System (ISIS). The R1200R grating and the Red+ CCD were used to obtain the red spectra centered on $\lambda=6560\,\mathrm{\AA}$ (2007 data) and on $\lambda=6521\,\mathrm{\AA}$ (2009 data). A slit width of 1.2\,arcsec was used for the 2007 observations and a 1\,arcsec slit for the 2009 ones.

The setup \textbf{WHT-B}lue denotes WHT data obtained using the ISIS spectrograph with the R600B grating and the blue EEV10. The grating was centered on $\lambda=4388\,\mathrm{\AA}$ with a 1\,arcsec slit (2006 observations), on $\lambda=4500\,\mathrm{\AA}$ with a 1.5\,arcsec slit (2007 data), $\lambda=4339\,\mathrm{\AA}$ with a slit width of 0.62\,arcsec (2008) and on $\lambda=4349\,\mathrm{\AA}$ with a 1.04\,arcsec slit during the 2009 observations.


\begin{table*}
\centering
\begin{minipage}{160mm}
\caption{Journal of observations. Observers: P. F. L. Maxted (P. M.), T. Augustijn (T. A.), T. R. Marsh (T. M.), Luisa Morales-Rueda (L. M.), G. Nelemans (G. N.),  
C. Copperwheat (C. C.),  
R.A. Wade (R. W.) and M. A. Stark (M. S.). 
}\label{tbl:obs}
\centering
\begin{tabular}{lccccc}\hline\hline
Date 			&	Setup	&	$\lambda$ region	&	 \# Spectra	&	Mean dispersion	&	Observer(s)		\\
				&			&					&			&	($\rm\AA/pixel$)		&				\\\hline
11 - 19/04/00		&	INT-R	&	$H_{\alpha}$		&	4		&	0.39				&	P. M.		\\
08 - 13/03/01		&	INT-R	&	$H_{\alpha}$		&	11		&	0.39				&	P. M.			\\
01 - 07/05/01		&	INT-R	&	$H_{\alpha}$		&	9		&	0.39				&	P. M.			\\
26 - 30/03/02		&	SAAO	&	Blue				&	6		&	0.49				&	T. M.			\\
25 - 27/04/02		&	INT-B	&	Blue				&	6		&	0.48				&	T. A. \& T. M.	\\
09 - 16/04/03		&	INT-B	&	Blue				&	27		&	0.48				&	T. M.			\\
30/03 - 05/04/04	&	SAAO	&	Blue				&	5		&	0.49				&	L. M.			\\
23 - 24/06/05		&	SAAO	&	Blue				&	2		&	0.50				&	L. M.			\\
06/02/06			&	WHT-R	&	$H_{\alpha}$		&	2		&	0.22				&	G. N.			\\
				&	WHT-B	&	Blue				&	2		&	0.44				&	G. N.			\\
09/03/07			&	WHT-B	&	Blue				&	4		&	0.44				&	service		\\
27/03 - 07/04/07	&	INT-B	&	Blue				&	11		&	0.48				&	T. M.			\\
29 - 31/03/07		&	WHT-R	&	$H_{\alpha}$		&	5		&	0.25				&	G. N.			\\
				&	WHT-B	&	Blue				&	5		&	0.44				&	G. N.			\\
21 - 22/03/08		&	INT-B	&	Blue				&	4		&	0.48				&	C. C.			\\
01/05/08			&	WHT-R	&	$H_{\alpha}$		&	2		&	0.49				&	P. M.			\\
				&	WHT-B	&	Blue				&	2		&	0.44				&	P. M.			\\
11/03/09			&	INT-B	&	Blue				&	2		&	0.48				&	C. C.			\\
30/04/09			&	WHT-R	&	$H_{\alpha}$		&	4		&	0.25				&	T. M.			\\
				&	WHT-B	&	Blue				&	4		&	0.44				&	T. M.			\\
03/04/10			&	NOT		&	$H_{\alpha}$ + Blue	&	1		&	0.03				&	service		\\
06/12/07 - 23/03/10	&	HET		&	$H_{\alpha}$ + Blue	&	7		&	0.12				&	R. W \& M. S.		\\\hline\hline
\end{tabular}
\end{minipage}
\end{table*}

The \textbf{NOT} data were taken using the 2.56\,m Nordic Optical Telescope. The FIES (Fibre-fed Echelle Spectrograph) highest resolution fiber (the 1.3 arcsec fibre offering a spectral resolution of R = 67000) covered the entire spectral range 3700 - 7300 $\mathrm{\AA}$ without gaps in a single fixed setting. 

The final setup, \textbf{HET}, refers to data from the bench-mounted echelle fibre-fed High Resolution Spectrograph, mounted on the 9.2m Hobby-Eberly Telescope operated in its R = 15000 resolution mode. In the ``2x3'' on-chip binning mode that was used, the dispersion was about 6 km/s per binned output pixel.  A 2-arcsecond optical fibre was used for the stellar target, and two additional fibers were used to record the sky spectrum.  A cross-dispersing grating with 600 grooves per millimeter was used, centering $\lambda \approx 5822\,\mathrm{\AA}$ at the boundary between the ``blue" and ``red" CCDs. The useful wavelength coverage extended from 4810$\,\mathrm{\AA}$ to 6760$\,\mathrm{\AA}$.


To reduce the spectra from the INT, WHT and SAAO, standard Starlink routines were used. Flatfields were taken to correct for the pixel to pixel variations in the CCD and the bias correction was carried out by using the overscan region in each CCD frame. The objects were extracted with the optimal extraction algorithm of \citet{mar1989}. CuAr+CuNe arc spectra were taken before and after each target spectrum or after each set of two spectra at the target's position to calibrate these in wavelength. Fourth order polynomials were computed to fit the lines in the arcs and the solutions were used for the calibration of the corresponding spectra.\\
The NOT data were reduced with the automatic data reduction software package FIEStool\footnote{Developed by Eric Stempels (http://www.not.iac.es/instru-ments/fies/fiestool/FIEStool.html)}, which makes use of the IRAF and NumArray packages via a Python interface. After preprocessing, the raw frames were debiased and subsequently divided by a 2D normalized flatfield, correcting for the shape of each spectral order. Next the science spectra were extracted using the optimal extraction algorithm from \citet{hor1986} and corrected for the blaze shape. The wavelength calibration was done from a ThAr lamp spectrum taken right before the science data.
For the HET spectra we used standard IRAF tasks, organized using``pipeline" scripts, to process the images and extract the spectra. Observations of PG1018-047 were taken as pairs of 750-second exposures.
Each pair was combined within IRAF, using the ``crreject" option in the task ``imcombine" to reduce cosmic-ray contamination; additional rejection of cosmic-ray artifacts was done later by hand. 
As a result of this observational effort, we have a total of 125 spectra of \pg. 

\section{Results}

\subsection{The optical spectrum of PG\,1018--047}


\begin{figure*}
\centering
\begin{minipage}{160mm}
\centering
\epsfig{file=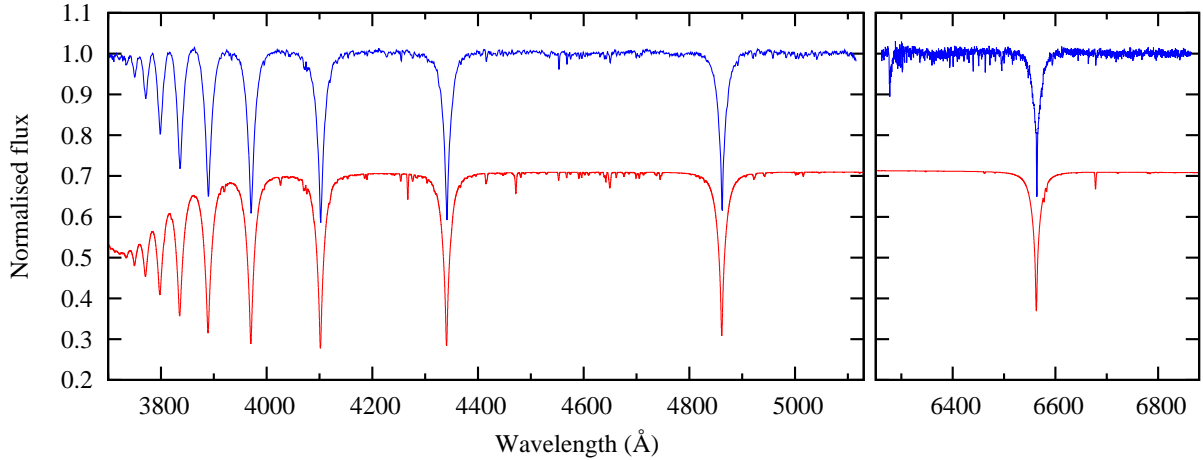, width=6.35cm, angle=270}

\caption{The mean spectrum of PG\,1018--047 for the region covering H$_\beta$ to H$_\iota$ (left) and the region around H$_\alpha$ (right). Plotted below the mean spectrum (shifted down for clarity) is a model spectrum for an sdB star (see text for details). Lines from the K-dwarf are clearly seen in the red part, but the lines in the blue are from the sdB. }\label{fig:optspec}
\end{minipage}
\end{figure*}


The normalised optical spectrum of PG\,1018--047 is shown in Figure \ref{fig:optspec}. It is obtained by averaging over a number of the ING/SAAO spectra taken at the same orbital phase, calculated using our best orbital solution from section \ref{sec:orbsol}. In order to normalise the spectra, we fitted third order polynomials to the regions free from absorption lines and divided the spectra by these fits.  
The mean spectrum shows several metal lines in the region between H$_\beta$ and H$_\gamma$. However, the \ion{He}{I} line at 4472\,\AA, which is typical of sdB stars is absent. Instead the \ion{Si}{iii} triplet (4553, 4568, 4575\,\AA) is the strongest feature. Several lines in the region between 4600 and 4700\,\AA\ can be identified with \ion{O}{ii} lines and possibly also \ion{N}{ii}.
We made a fit to the mean spectrum of PG\,1018--047 using the LTE model grids of Heber et al.~(2000), with explicit metals of solar composition and abundances depleted by 0.0, 0.5, 1.0, 1.5 and 2.0 dex relative to solar. A reasonable fit is achieved with $T_{\rm eff}$\,=\,30500\,$\pm$\,200 K, log\,$g$\,=\,5.50\,$\pm$\,0.02, and with the N and O abundance 1/10 of the solar value.  
In Figure~1, a synthetic spectrum with these parameters broadened to match the resolution of the observed spectrum is shown. The model spectrum contains helium at a fraction log\,N(He)/N(H) = -3.0, but even this is clearly too much. In order to make helium fit with the observed spectrum, the model must be depleted to log\,N(He)/N(H) $<$ -\,4. It is also clear that the abundances of the various elements are quite far from solar composition relative to each other, which is not unusual for the sdBs (Heber et al.~2000).
Note that the K-star contributes some light also in the blue part of the spectrum, but insufficient to make any lines clearly visible in the spectrum. However, the contribution to the continuum might still be sufficient to affect the fitting procedure. In the H$_\alpha$ region, however, metal lines from the K-star are clearly seen. Our high S/N mean spectrum has too low resolution to reliably infer the abundances of the individual components, and our high resolution spectra have insufficient S/N. 


\begin{table*}
\centering
\begin{minipage}{160mm}
\caption{Narrow absorption lines present in the optical spectrum (Figure \ref{fig:optspec}). The identification was done using the mid- and high-resolution spectral library of \citet{mon1997} and also \citet{ral2010}. The lines with a * were included in the template to determine the radial velocities of the secondary (see section \ref{sec:sec}). For reference, we have added the wavelengths for the Balmer lines we adopted in our analysis as well.}\label{tbl:seclines}
\centering
\begin{tabular}{clclclcl}
\hline\hline
$\lambda$ (\AA) & Element & $\lambda$ (\AA) & Element & $\lambda$ (\AA) & Element & $\lambda$ (\AA) & Element \\\hline
4153.30 & \ion{O}{ii}				& 4596.17 	& \ion{O}{ii}			& 6337.28*   	& & 6609.05*	&				\\
4164.79 &	 \ion{Fe}{iii}, blend		& 4630.54 	& \ion{N}{ii}			& 6359.45*   	& & 6614.42*	&				\\
4189.80 & \ion{O}{ii}				& 4639.70 	& \ion{O}{ii}, blend		& 6363.71*	& & 6637.09*	&				\\
4253.59 & \ion{S}{iii} + \ion{O}{ii} 	& 4642.26\,\, 	& \ion{O}{ii}, \ion{N}{ii}	& 6394.45*   	& & 6644.39*	&				\\
4276.74 & \ion{O}{ii}, blend 		& 4649.14\,\, 	& \ion{O}{ii} 			& 6400.43*   	& & 6664.37*	&				\\
4414.91 & \ion{O}{ii} 			& 4661.04\,\, 	& \ion{O}{ii} 			& 6408.96*   	& & 6678.97*	&				\\
4416.58 & \ion{O}{ii}				& 4676.23\,\, 	& \ion{O}{ii}			& 6412.37* 	& &			&				\\
4442.49 & \ion{O}{ii} 			& 4700.31\,\, 	& \ion{O}{ii}, blend		& 6422.05* 	& &  4101.735 	& $H_{\delta}$ 		\\
4552.65 & \ion{Si}{iii}  			& 4705.44\,\, 	& \ion{O}{ii}			& 6431.44* 	& & 4340.465 	& $H_{\gamma}$ 	\\ 
4567.87 & \ion{Si}{iii} 			& 4710.04\,\, 	& \ion{O}{ii}			& 6439.70* 	& &  4861.327 	& H$_{\beta}$ 		\\
4574.78 & \ion{Si}{iii} 			&			&					& 6450.51* 	& & 6562.800 	& H$_{\alpha}$ 	\\
4590.57 & \ion{O}{ii}				& 6335.83*	&  					& 6456.49* 	& &			&				\\
\hline\hline
\end{tabular}
\end{minipage}
\end{table*}

\subsection{The orbit of PG\,1018--047}\label{sec:orbsol}

\subsubsection{Radial velocity measurements}\label{sec:radvelmeas}

The radial velocities (RVs) of the INT, WHT, SAAO and NOT spectra were determined following the procedure described by \citet{mor2003}, i.e. least squares fitting of a line profile model. This line profile model was built up from three Gaussians per Balmer line with different widths and depths, but with a common central wavelength position which varies between the spectra. The parameters of the Gaussians were optimized by comparing the model to the normalized average spectrum over all observations; see \citet{mmm2000c} for further details of this procedure. For the blue spectra we fit simultaneously for \Hb, \Hg\ and \Hd, whereas for the red spectra only the \Ha\ line can be fitted. The RV of the NOT spectrum was determined using a model containing \Ha, \Hb\ and \Hg\ (see also Table \ref{tbl:seclines}). The radial velocities from the 7 HET spectra are obtained from the \Hb\ absorption line using simple Gaussian fitting to the core of the line within the IRAF "splot" task. 
A list of the radial velocitities and the uncertainties measured is given in Table \ref{tbl:radvel}. 


\begin{table*}
\centering
\begin{minipage}{160mm}
\caption{The 125 radial velocity measurements for PG\,1018--047 with their formal errors from the least squares fitting routine. }\label{tbl:radvel}
\centering
\begin{threeparttable}
\begin{tabular*}{14cm}{p{2cm}p{2cm}p{2cm}p{2cm}p{2cm}p{2cm}}\hline\hline
HJD			&	RV		&	HJD		&	RV		&	HJD		&	RV		\\
-2450000		&	(km/s)	&	-2450000	&	(km/s)	&	-2450000	&	(km/s)	\\\hline
1646.47135	& 31.3 $\pm$ 4.4 &2741.52795	& 49.8 $\pm$ 3.0 &4189.54710	& 52.1 $\pm$ 1.4 \\
1646.47650	& 20.5 $\pm$ 4.3 &2741.53909	& 53.3 $\pm$ 3.1 &4189.54712	& 42.9 $\pm$ 1.0 \\
1654.46074	& 24.7 $\pm$ 3.8 &2741.54857	& 48.8 $\pm$ 3.4 &4190.40980	& 51.9 $\pm$ 1.0 \\
1654.46780	& 32.1 $\pm$ 3.7 &2741.55807	& 48.2 $\pm$ 4.3 &4190.40982	& 46.0 $\pm$ 0.7 \\
1977.48220	& 49.6 $\pm$ 5.3 &2743.42040	& 50.4 $\pm$ 4.5 &4190.42420	& 53.5 $\pm$ 1.0 \\
1978.62533	& 55.2 $\pm$ 3.8 &2743.43451	& 60.3 $\pm$ 4.5 &4190.42422	& 50.7 $\pm$ 0.7 \\
1979.54044	& 45.0 $\pm$ 3.2 &2743.52128	& 52.2 $\pm$ 6.3 &4191.59677	& 52.1 $\pm$ 1.4 \\
1979.59958	& 65.5 $\pm$ 3.3 &2744.35895	& 54.3 $\pm$ 3.5 &4191.59677	& 44.1 $\pm$ 1.0 \\
1979.60655	& 54.0 $\pm$ 3.3 &2744.37308	& 49.7 $\pm$ 2.8 &4193.37192	& 51.4 $\pm$ 7.8 \\
1982.51489	& 52.2 $\pm$ 3.9 &2744.39186	& 49.4 $\pm$ 2.8 &4195.35718	& 41.6 $\pm$ 1.9 \\
1982.52185	& 54.6 $\pm$ 4.1 &2744.40598	& 51.4 $\pm$ 2.7 &4195.37121	& 36.4 $\pm$ 1.7 \\
1982.55421	& 57.8 $\pm$ 4.0 &2744.42529	& 53.9 $\pm$ 2.6 &4195.61447\tnote{$\spadesuit$}	& 35.5 $\pm$ 2.5 \\
1982.56118	& 52.6 $\pm$ 3.9 &2744.43941	& 52.2 $\pm$ 2.6 &4195.62457\tnote{$\spadesuit$}	& 38.1 $\pm$ 4.5 \\
1982.60059	& 57.5 $\pm$ 3.9 &2745.37982	& 48.8 $\pm$ 3.5 &4197.51456	& 43.8 $\pm$ 1.8 \\
1982.60755	& 53.2 $\pm$ 4.0 &2745.39394	& 45.3 $\pm$ 3.1 &4197.52858	& 43.6 $\pm$ 1.9 \\
2031.36769	& 62.1 $\pm$ 8.3 &2745.40987	& 51.4 $\pm$ 2.8 &4441.00811 	& 39.7 $\pm$ 2.0 \\  
2031.37466	& 41.4 $\pm$ 8.4 &2745.42399	& 51.9 $\pm$ 2.7 &4469.93158 	& 36.0 $\pm$ 2.0 \\ 
2031.38641	& 49.5 $\pm$ 4.4 &2746.44131	& 54.6 $\pm$ 2.5 &4502.85094 	& 33.5 $\pm$ 2.0 \\
2032.48243	& 54.9 $\pm$ 3.0 &2746.45544	& 51.0 $\pm$ 2.5 &4547.37872	& 33.0 $\pm$ 1.7 \\
2032.49633	& 53.6 $\pm$ 3.5 &2746.47412	& 55.4 $\pm$ 2.2 &4547.40665	& 31.1 $\pm$ 1.6 \\
2033.39672	& 53.8 $\pm$ 4.1 &2746.48824	& 44.8 $\pm$ 2.3 &4547.62144	& 29.6 $\pm$ 2.0 \\
2033.40369	& 53.8 $\pm$ 4.3 &3095.38227	& 26.8 $\pm$ 6.7 &4547.64844	& 32.4 $\pm$ 2.2 \\
2037.45584	& 54.1 $\pm$ 3.8 &3095.40336	& 19.0 $\pm$ 6.6 &4550.71405 	& 26.7 $\pm$ 2.0 \\
2037.46282	& 50.9 $\pm$ 3.5 &3098.33704	& 12.9 $\pm$ 9.8 &4562.68846 	& 29.6 $\pm$ 2.0 \\
2360.36683\tnote{*}	& -33.8 $\pm$ 25.0 &3101.31890	& 18.8 $\pm$ 9.2 &4588.35123	& 29.7 $\pm$ 2.5 \\
2360.37745\tnote{*}	& -4.6 $\pm$ 22.9 &3101.32956	& 19.7 $\pm$ 8.7 &4588.35125\tnote{$\clubsuit$}	& 34.2 $\pm$ 2.6 \\
2362.36641\tnote{*}	& 17.5 $\pm$ 19.3 &3545.21953	& 51.4 $\pm$ 7.2 &4588.35646\tnote{$\clubsuit$}	& 22.8 $\pm$ 2.6 \\
2362.38049\tnote{*}	& 5.4 $\pm$ 20.3 &3546.20756	& 52.6 $\pm$ 5.8 &4588.35648	& 29.5 $\pm$ 2.4 \\
2364.39275	& 23.1 $\pm$ 6.0 &3772.55466	& 35.6 $\pm$ 0.8 &4902.42320	& 34.8 $\pm$ 4.5 \\
2364.40338	& 17.9 $\pm$ 6.7 &3772.55467	& 26.7 $\pm$ 1.2 &4902.44418	& 36.6 $\pm$ 4.9 \\
2390.51113	& 34.6 $\pm$ 17.1 &3772.68564\tnote{$\bullet $}	& 15.3 $\pm$ 0.9 & 4952.38360	& 55.2 $\pm$ 2.8 \\
2390.52179	& 32.8 $\pm$ 9.0 &3772.68565	& 34.6 $\pm$ 1.4 &4952.38359	& 54.3 $\pm$ 5.1 \\
2391.37270	& 30.1 $\pm$ 4.3 &4169.43214	& 43.8 $\pm$ 1.2 &4952.39761	& 53.6 $\pm$ 3.9 \\
2391.38337	& 32.7 $\pm$ 3.6 &4169.44290	& 47.3 $\pm$ 1.4 &4952.39762	& 49.6 $\pm$ 2.2 \\
2392.36580	& 28.3 $\pm$ 2.1 &4169.45704	& 56.5 $\pm$ 2.0 &4952.41400	& 58.4 $\pm$ 1.8 \\
2392.37993	& 24.5 $\pm$ 2.0 &4169.46774	& 52.6 $\pm$ 6.1 &4952.41401	& 51.8 $\pm$ 2.9 \\
2739.51074	& 45.0 $\pm$ 3.6 &4186.54793	& 38.0 $\pm$ 3.6 &4952.42802	& 51.1 $\pm$ 2.3 \\
2739.52029	& 46.7 $\pm$ 3.8 &4186.55810	& 40.1 $\pm$ 11.8 &4952.42802	& 58.4 $\pm$ 1.5 \\
2740.44514	& 49.8 $\pm$ 3.2 &4188.41925	& 39.9 $\pm$ 3.0 &5202.92761 	& 36.5 $\pm$ 2.0 \\
2740.45465	& 51.9 $\pm$ 3.0 &4188.43328	& 42.1 $\pm$ 2.9 &5278.71693 	& 32.8 $\pm$ 2.0 \\
2741.50896	& 49.7 $\pm$ 3.1 &4189.53608	& 49.9 $\pm$ 1.5 &5295.96094 	& 35.0 $\pm$1.6  \\ 
2741.51846	& 49.4 $\pm$ 3.1 &4189.53609	& 42.6 $\pm$ 1.0 & & \\\hline\hline
\end{tabular*}
\begin{tablenotes}
\item[*] \small{The SAAO science frames belonging to these 4 radial velocities were made with very marginal observing conditions. The resulting unreliable RVs are therefore not considered in the remaining analysis.}
\item[$\bullet$] \small{This is a discrepant WHT data point, taken during service observations in 2006. It is not in line with the other data taken at on the same night and therefore given no weight in the remaining analysis.}
\item[$\spadesuit$] \small{For the blue WHT science frames belonging to these two RVs, only a single arc frame was available, taken before the first of the two subsequent observations.}
\item[$\clubsuit$] \small{These two WHT observations in the $H_{\beta}$ - $H_{\delta}$ region had no flat fields or bias frames available.}
\end{tablenotes}
\end{threeparttable}
\end{minipage}
\end{table*}

\subsubsection{Orbital parameters}\label{sec:orbpar}


Once the radial velocities for all spectra were known, we used the floating mean periodogram \citep{CMB1999}, a generalization of the well-known Lomb-Scargle periodogram \citep{lom1976, sca1982}, to determine the most probable frequencies (periods) present in the data. The method consists in fitting the radial velocity data with a model composed of a sinusiod plus a constant of the form: 
\begin{equation}
v = \gamma + K\sin[2\pi f(t-t_{0})],\label{eq:orbit}
\end{equation}
with $f$ the frequency and $t$ the time of observation. This means we assume the binary system to have a circular orbit with semi-amplitude $K$ and a systemic velocity $\gamma$. For each frequency ($f$=1/$P$) we perform least squares fitting of the data, solving for $\gamma$ and $K$ simultaneously using singular value decomposition \citep{pre2002}. In this way we can obtain the $\chi^{2}$ statistic of the model as a function of frequency, or in other words the periodogram.


In our initial period determination, the $\chi^{2}$ values turned out larger than expected given the number of data points, indicating that there must have been an extra unaccounted source of uncertainty, most likely due to systemic effects, intrinsic variability of the star or slit-filling errors (see also \citealt{mor2003}). Such errors are unlikely to be correlated with either the orbit or the statistical errors we have estimated. To allow for this we compute the level of systematic uncertainty ($\sigma$) per telescope that when added in quadrature to our error estimates gives a reduced  $\chi^{2}\approx 1$ for each of the INT, WHT subsets, telescope by telescope, relative to the preliminary fit. 
We increased the errors by $\sigma_{\rm{INT}}=\,$2.15 km s$^{-1}$ and $\sigma_{\rm{WHT}}=\,$3.4 km s$^{-1}$ for respectively the INT and WHT data. The formal NOT and HET errors were left unchanged since these are fibre-fed instruments which do not suffer from normal slit-guiding errors. Also the 9 SAAO RVs were left as they were. Second we used the average residual from each data set to our best orbital fit at that point to apply offsets to the INT, WHT and SAAO data sets\footnote{Table \ref{tbl:fitrvs} presents the RVs at this stage.} (respectively 0.85, -2.00 and 4.63 km s$^{-1}$, predicted minus observed). Finally, we scale the errors of the entire data set multiplicatively by a factor 1.244 to obtain a $\chi^{2}$ value equal to the degrees of freedom (dof). 

\begin{figure}
\begin{minipage}{85mm}
\epsfig{file=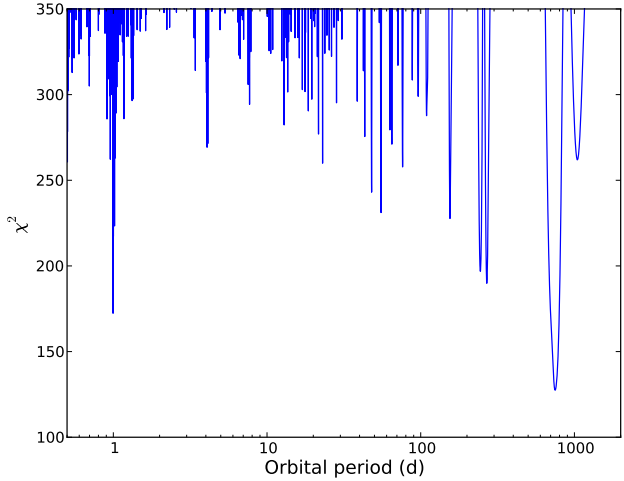, width=9.5cm}
\caption{The radial velocity periodogram for PG\,1018--047, showing the most probable orbital periods present in the data.}\label{fig:oripgram}
\end{minipage}
\end{figure}

Figure \ref{fig:oripgram} shows the resulting radial velocity periodogram for \pg\ in the region where we find the lowest $\chi^{2}$ values. The best solution is found around 760 days, followed by a group of 1 day aliases and the yearly aliases of the long period around 250 days. An overview of the best aliases with their corresponding $\chi^{2}$ values is given in Table \ref{tbl:bestaliases}. Given the large inhomogeneity of the data sets, we realize that the treatment of the errors described above might not be perfect in all details. Therefore we have considered alternative methods to weight the data and errors as well, but the essential result, the long period, was unchanged throughout. \\


\begin{table}
\centering
\begin{minipage}{80mm}
\caption{The best orbital periods found from the radial velocity periodogram for PG\,1018-047. The $\chi^{2}$ value, mass function of the companion and the minimum mass of the companion obtained by assuming an sdB mass of 0.5$\,\mathrm{M_{\odot}}$ are also given.}\label{tbl:bestaliases}
\centering
\begin{tabular}{ccccc}\hline\hline
Alias	&	Period (d)	&	$\chi^{2}$	&	$f_{m}$ ($M_{\odot}$)	&	$M_{2min}$ ($M_{\odot}$)	\\\hline
1	& 759.80		& 117		& 0.17245					&  0.58927					\\
2 	& 0.9987		& 188		& 0.00018					&  0.03704					\\
3 	& 241.35		& 238		& 0.02602 				&  0.24313					\\
4 	& 267.93		& 258		& 0.05582 				&  0.34031					\\
5	& 1.0184		& 259		& 0.00013					&  0.03332					\\\hline\hline

\end{tabular}
\end{minipage}
\end{table}

In Table \ref{tbl:bestaliases} we also quote the mass functions of the companion, calculated using
\begin{equation}\label{eq:minmass}
f_{m}=\frac{M_{MS}^{3}\sin^{3}i}{(M_{sdB}+M_{MS})^{2}}=\frac{PK_{sdB}^{3}}{2\pi G}.
\end{equation}
The minimum mass of the companion, assuming a typical  sdB mass of $0.5\rm\,M_{\odot}$ \citep{heb1984}, is also given in each case. These numbers will be used later to constrain the nature of the companion star and vice versa to show that our orbital solution is plausible given the properties of the secondary.

\begin{figure*}
\centering
\begin{minipage}{160mm}
\begin{center}
\epsfig{file=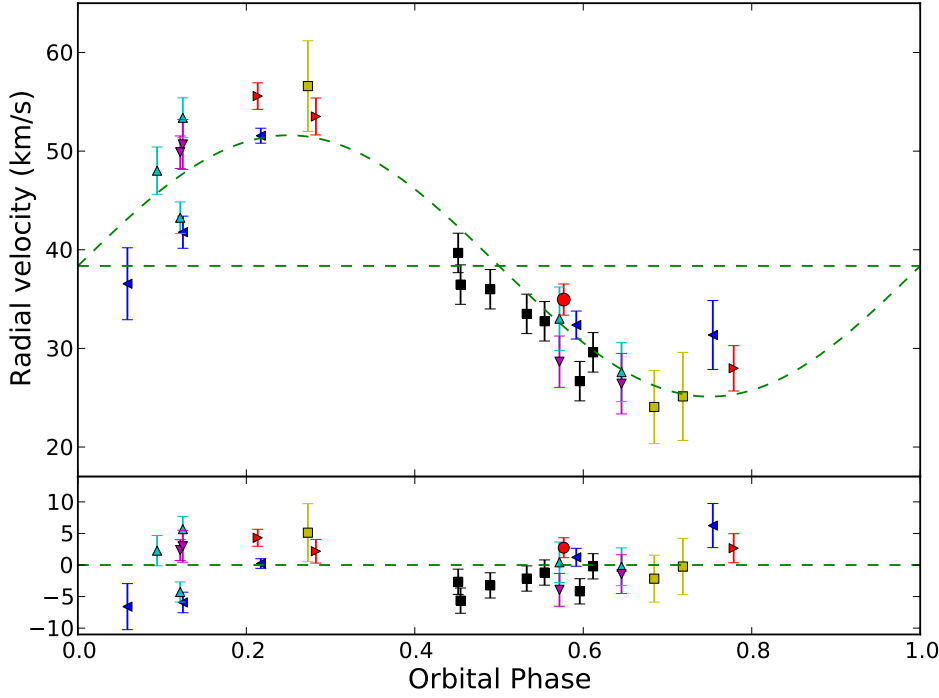, width=14cm}
\caption{Radial velocity curve for the sdB component of \pg. We have averaged the RVs per observation run. The small lower panel shows the residuals. The black squares are the 7 RVs from the HET spectra, the red dot is the NOT/FIES data point, yellow squares are the SAAO observations, the blue left triangles and red right triangles are the respectively blue and red INT data points, and the cyan up and magenta down triangles correspond to the blue and red WHT radial velocities.}\label{fig:bestorbsol}
\end{center}
\end{minipage}
\end{figure*}

Table \ref{tbl:bestorbsol} lists the orbital parameters for our best orbital solution found for \pg\ and in Figure \ref{fig:bestorbsol} the radial velocity curve folded on the period is shown. 


\begin{table}
\centering
\begin{minipage}{80mm}
\caption{The best orbital solution for \pg\, assuming a circular and eccentric orbit.}\label{tbl:bestorbsol}
\centering
\begin{tabular}{lrr}\hline\hline
						&	Circular				& Eccentric			\\\hline
$\mathrm{P_{orb}}$ (d)		&	759.8$\pm$5.8			& 755.9$\pm$5.1		\\
$\gamma$ (km/s)			&	38.2$\pm$0.5			& 38.0$\pm$0.9		\\
$e$						&	0					& 0.246$\pm$0.052		\\
$\omega$	 ($^{\circ}$)		&						& 0$\pm$24			\\
HJD$_{0}$ (d)				&	2453335.0$\pm$10.5	& 2453343.0$\pm$14.7	\\
$\mathrm{K_{sdB}}$ (km/s)	&	13.0$\pm$0.8			& 12.6$\pm$0.8		\\
$\mathrm{K_{MS}}$ (km/s)	&	8.1$\pm$1.0			&					\\\hline\hline
\end{tabular}
\end{minipage}
\end{table} 



\subsubsection{The secondary orbit}\label{sec:sec}

Narrow metal lines of a cool companion were present in addition to the Balmer absorption lines from the sdB star, allowing us to obtain radial velocity variations for the secondary star. The blue spectra did not contain enough signal from the companion to be useful, so we focussed on the HET, NOT and red WHT/INT spectra.
We determined the radial velocities of the secondary by means of cross correlation with a template spectrum \citep{td1979}. 

For the HET data we cross-correlated with 61 Cyg B (K7V), which was obtained as part of a different program on April 15, 2010, using HRS in its R = 30,000 mode with a 316 groove/millimeter cross disperser centred at 6948\,\AA. As a result only seven orders from the red CCD (six, on one night) were available for cross correlation, covering 6000 - 6760\,\AA\ (orders containing \Ha\ or strong telluric lines were excluded).

For the other data we constructed a template from our data using the average red spectrum of PG\,1018--047 (the upper panel in Figure \ref{fig:optspec}). Every (by eye) recognisable absorption feature of the secondary was fitted with a Gaussian profile, similar to the method described in section \ref{sec:radvelmeas}. All features possibly originating in the primary or of telluric origin were masked out, leaving a total of 19 lines from the cool companion in the template (see also Table \ref{tbl:seclines}).

Unfortunately the quality of the secondary radial velocities obtained from the red INT and WHT spectra were not sufficient to derive any trustworthy orbital period from a periodogram or to estimate the orbital semi-amplitude $K_{MS}$ for the secondary. We therefore used our best orbital period obtained from the primary RVs (Table \ref{tbl:bestorbsol}) to phase-bin the 37 medium-resolution red spectra before the cross-correlation routine. In total 7 out of 40 bins are filled. The HET and NOT spectra were left unbinned.


\begin{table}
\centering
\begin{minipage}{80mm}
\caption{The radial velocities of the cool companion.}\label{tbl:secrv}
\centering
\begin{tabular}{cccc}\hline\hline
				&	\# spectra	&	Average		&	RV			\\
Bin				&	in bin	&	phase		&	(km/s)		\\\hline
0.100 - 0.125		&	2		&	0.1247		&	35.7 $\pm$ 1.9	\\
0.125 - 0.150		&	7		&	0.1277		&	32.9 $\pm$ 1.3	\\
0.200 - 0.225		&	11		&	0.2176		&	29.2 $\pm$ 3.3	\\
0.250 - 0.275		&	9		&	0.2867		&	29.0 $\pm$ 4.3	\\
0.4556			&	HET		&				&	36.7 $\pm$ 2.8 \\
0.4585			&	HET		&				&	34.8 $\pm$ 0.9 \\
0.4937			&	HET		&				&	39.0 $\pm$ 1.5 \\
0.5370			&	HET		&				&	41.3 $\pm$ 1.1 \\
0.5583			&	HET		&				&	41.9 $\pm$ 0.7 \\
0.575 - 0.600		&	2		&	0.5759		&	43.5 $\pm$ 2.2 \\
0.5810			&	NOT		&				&	40.2 $\pm$ 0.6 \\
0.6000			&	HET		&				&	44.5 $\pm$ 1.1 \\
0.6158			&	HET		&				&	43.6 $\pm$ 0.6 \\
0.625 - 0.650		&	2		&	0.6496		&	39.4 $\pm$ 4.9 \\
0.775 - 0.800		&	4		&	0.7826		&	44.8 $\pm$ 4.9	\\\hline\hline
\end{tabular}
\end{minipage}
\end{table}

\begin{figure*}
\centering
\begin{minipage}{160mm}
\begin{center}
\epsfig{file=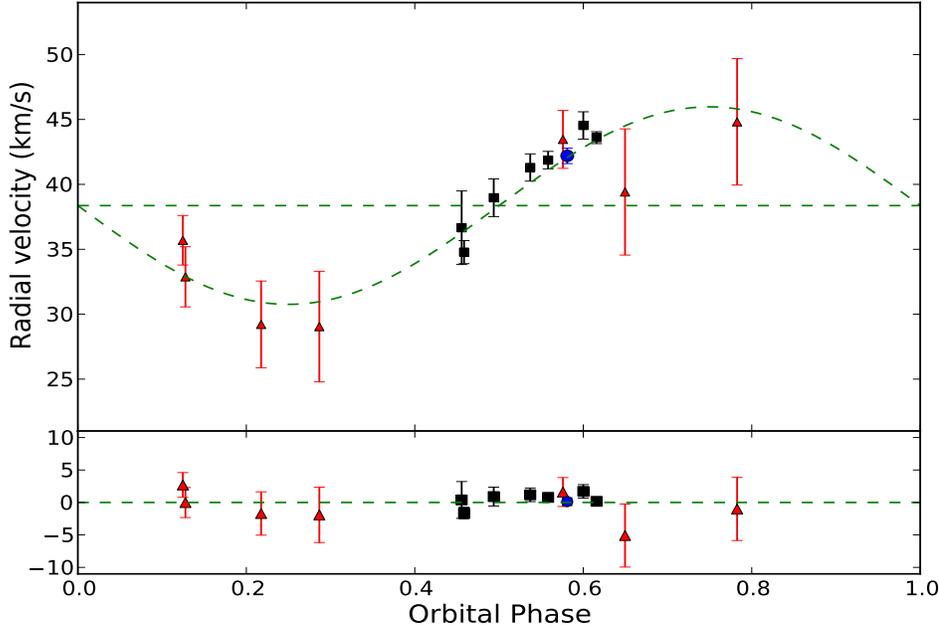, width=14cm, height=9.5cm}
\caption{The orbital fit for the secondary star. The red triangles are the radial velocities from the cool companion, obtained from the intermediate resolution spectroscopy. The NOT measurement is plotted with a blue dot and the HET data with black squares. The radial velocity curve (dashed green) is also shown.}\label{fig:secfit}
\end{center}
\end{minipage}
\end{figure*}

In Figure \ref{fig:secfit} we have plotted phase versus radial velocity. Using least squares fitting we estimated the projected semi-amplitude of the secondary to be $K_{MS} = 8.1 \pm 1.0$ km/s,
which constrains the mass ratio of the secondary to the primary star to be $q = M_{MS}/M_{sdB} = 1.6 \pm 0.2$. The $\chi^{2}$ value of the fit was 13.29, given 15 data points minus 2 fitting parameters.

\subsubsection{Eccentricity of the orbit}

In sections \ref{sec:orbpar} and \ref{sec:sec} the analysis was based upon the assumption that PG1018-047 has a circular orbit. However, we also tried fitting eccentric orbits using the Markov chain Monte Carlo (MCMC; \citealt{GRS1995}) method obtaining an eccentricity of $0.246 \pm 0.052$; the favoured period for these fits decreased from $759.8 \pm 5.7$ days to $755.9 \pm 5.1$ days and had $\chi^{2} = 152$, starting from the radial velocities in Table \ref{tbl:fitrvs}. To obtain $\chi^{2}/\rm{dof}  = 1$ we needed to scale the errors by a factor 1.153. Table \ref{tbl:bestorbsol} presents the orbital parameters. The argument of periapsis ($\omega$) is defined as the angle in the orbital plane between the ascending node and the line of apsides. 0 degrees indicates that the major axis of the ellipse is in the plane of sky. Note as well that HJD$_{0}$ is given as the ascending node passage minus $\mathrm{P_{orb}/4}$ to be consistent with our definition for the circular orbit (see eq. (\ref{eq:orbit})).

Computing the F-statistic we tested whether eccentricity was needed in our model. Assuming a locally linear model, the variable $X = \chi^{2}_{circ} - \chi^{2}_{ecc}$ should itself have a $\chi^2$ distribution with 2 dof (the eccentricity fit parameters), independent of the $\chi^{2}$ of the 114 degrees of freedom of the eccentric orbit. 
\begin{equation*}
\rm{F} = \frac{(\chi^{2}_{circ} - \chi^{2}_{ecc})/2}{\chi^{2}_{ecc}/114} = 11.24
\end{equation*}
should then be distributed as F(2;114) under the null hypothesis that the orbit is circular. The chances of such a large value are very small, meaning we reject the null hypothesis at the 99.9\% significance level in favour of an eccentric orbit.
Although formally significant the heterogeneous nature of our data and the limited phase coverage lead us to be wary of claiming a definitive detection of eccentricity. However, it is equally true that the orbit could be significantly non-circular.

\subsection{The nature of the companion}

We determined the spectral type of the companion star by minimizing the residuals after subtracting different template star spectra from the mean red \pg\ spectrum in the wavelength region between 6390\,\AA\ and 6700\,\AA. We masked out \Ha, because the line contains a large contribution from the sdB. This optimal subtraction routine \citep{MRW1994} is sensitive to both the rotational broadening $v_{rot}\sin i$ and the fractional contribution $f$ of the companion star to the total flux. 
We used template stars from two different origins in the spectral range F0 to M9: Kurucz models (Munari et al. 2005, solar composition) and 43 real star templates \citep{mon1997}. All templates were prepared so that they had the same wavelength coverage and resolution as our normalised \pg\ spectrum in the \Ha\ region.

\begin{figure}
\centering
\begin{minipage}{80mm}
\epsfig{file=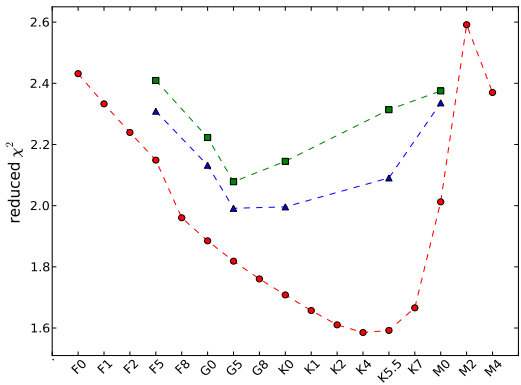, width=8cm, height=5.8cm}
\caption{Results of the optimal subtraction routine for the Kurucz templates. The circle curve are the results for dwarf stars, whereas the triangles and squares plot the reduced $\chi^{2}$ versus spectral type for respectively subgiant and giant stars.}\label{fig:kurucz}
\end{minipage}
\end{figure}

Figure \ref{fig:kurucz} clearly shows the difference in $\chi^{2}$ between the dwarf and (sub)giant templates. The main sequence models, which have the lowest $\chi^{2}$-values, show a distinct minimum at spectral type K4 - K6, whereas the (sub)giant models seem to converge to a spectral type G5. This is consistent with the set of template spectra from \citet{mon1997} as well, where we find the best $\chi^{2}$ for the K5V star HD\,201091 (61 Cyg A), followed by a K7V and K3V star. For M stars the worst values are found. The best non-dwarf templates were both G5 stars. 


Since \pg\ was observed by 2MASS, we adopt a similar approach as \citet{SW2003} as a complementary way to identify the spectral type of the companion star. 
We choose to focus on $J-K_{S}$ versus $B-V$ and $J-K_{S}$ versus $V-K_{S}$ and converted the Str\"omgren magnitudes to the Johnson system following the approach of \citet{tur1990}. The calculated colours are given in the lower part of table \ref{tbl:phot}. 


\begin{table}
\centering
\begin{minipage}{80mm}
\caption[Magnitudes and colours for PG\,1018--047.]{\textit{Magnitudes and colours for PG\,1018--047. The $B-V$ colour is calculated using the transformation formula from Turner (1990)}.}\label{tbl:phot}
\centering
\begin{tabular}{cccc}\hline\hline
\multicolumn{2}{c}{2MASS (infrared)}								&	\multicolumn{2}{c}{Str\"omgren (visual)}					\\
\multicolumn{2}{c}{(\underline{Skrutskie et al. 2006})}		&	\multicolumn{2}{c}{(\underline{Wesemael et al 1992})}		\\
J $\;\:$	&	= 13.298 (.026)								&	y 		&	= 13.320 (.005)					\\
H$\;\:$	&	= 12.980 (.027)								&	b-y 		&	= -0.086 (.004)					\\
K$_S$	&	= 12.928 (.033)								&	u-b		&	= -0.073 (.006)					\\
		&											&	m$_1$	&	= $\;$0.076 (.013)					\\\hline
		&	\multicolumn{2}{c}{ \underline{Calculated colours}}				&		\\
		&	 										B-V$\;$	&	 = -0.20 (.007)\\
		&											 J-K$_S$	&	 =	0.370 (.042)					\\
		&											 V-K$_S$	&	 = 0.392 (.038)					 \\\hline\hline
\end{tabular}
\end{minipage}
\end{table}


\begin{figure*}
\centering
\begin{minipage}{160mm}
\centering
\epsfig{file=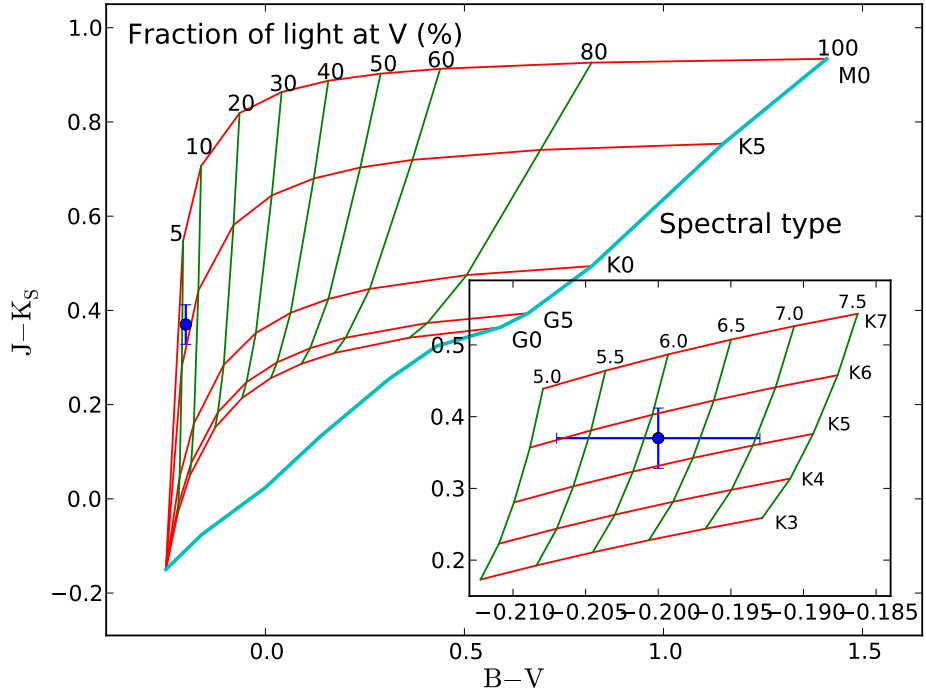, width=14cm}
\caption{Grid of composite colours in $B-V$ versus $J-K_{S}$ space by combining the light from a typical hot subdwarf with that of a population I main-sequence star assuming various fractional contributions to the total light in the $V$ band by the late-type star. The blue dot marks the values for PG\,1018--047. A close-up is shown in the small inset figure. The diagonal cyan line indicates the location of the population I main sequence.}
\label{fig:45}
\end{minipage}
\end{figure*}

The next step is to compare the \pg\ colours to a theoretical grid of colours for sdB-MS binaries while varying the fraction of light at $V$ that arises from the companion. Defining this fraction as
\begin{equation*}
f=\frac{F_{V_{\rm{MS}}}}{F_{V}}=\frac{F_{V_{\rm{MS}}}}{F_{V_{\rm{sdB}}}+F_{V_{\rm{MS}}}}, 
\end{equation*}
we find the combined colour (e.g. for B--V) from the expression
\small{}
\begin{eqnarray*}
\lefteqn{(B-V)_{\rm{sdB+MS}} = }\\
& & -2.5\log\left[(1-f)\cdot10^{-(B-V)_{\rm{sdB}}/2.5} + f\cdot 10^{-(B-V)_{\rm{MS}}/2.5}\right].
\end{eqnarray*}
\normalsize{}
Colour indices for late-type stars are taken from \citet{john1966}. The typical colours for a single sdB we take from \citet{SW2003}.

Figure \ref{fig:45} shows that as the fractional contribution $f$ at $V$ from the secondary increases, the $B-V$ index shifts to redder values. On the other hand, the cooler the companion becomes, the more the $K_{S}$-band dominates the $J-K_{S}$ index. Swapping the $B-V$ index for the $V-K_{S}$ colour shows a similar trend. Zooming in on PG\,1018--047, we find a spectral classification $K3 - K6$, which is consistent with the results from the optimal subtraction routine. We estimate the contribution of the secondary to be $6.1\pm1.0\,\%$ in the V-band. \\

\section{Discussion}\label{sec:discussion}

The role of the long period binary system \pg\ in refining our understanding of the origin and evolution of hot subdwarf stars depends on whether its present orbit is circular or eccentric.

\subsection{Evolution assuming a circular orbit}

If \pg\, indeed has a circular orbit, one can assume that tidal interaction has occurred between the sdB and the dK5 star. This means that from a theoretical point of view \pg\ might be an important system, as it becomes a good candidate for formation through the first stable Roche lobe overflow channel described by \citet{han2003}. From Figure 15 in \citet{han2003} we deduce that an sdB binary with a mid-K companion is feasible. 
However these systems are all subgiants and giants evolved from B stars (i.e. more massive stars at a later evolutionary state). 
Also, to have stable RLOF on the first giant branch onto a companion that is only 0.6-0.7$\,\rm{M_{\odot}}$ and still to end up with such a long period, mass transfer would have to be close to conservative to avoid excessive angular momentum loss. If we want the mass donor to evolve in a Hubble time, it must have started with an initial mass of 0.8$\,M_{\odot}$ or greater, meaning that the initial mass of the present K star must have been no more than $\sim$0.3-0.4$\,\rm{M_{\odot}}$. But then the initial mass ratio $q_0 \gtrsim 2$ is not compatible with conservative mass transfer on the red giant branch, leading to a contradiction. 

Second we can consider the alternative common-envelope prescription from \citet{nel2000}. In \citet{nel2010}, a population of sdB stars is simulated in which the first phase of mass transfer can be described by this alternative common-envelope prescription \citep{nel2001a}. Interestingly, in that model a substantial fraction of the sdB stars with low and intermediate mass main sequence companions have rather large orbital periods (100\,-\,1000 days, see their Figure 2) and the parameters of \pg\ actually fall right in a densely populated area of the model. Thus a sample of these long-period sdB binaries will give another way of testing the outcome of the common-envelope phase.

\subsection{Evolution assuming an eccentric orbit}

\citet{CW2011} propose that binaries similar to \pg\ can be the remnants of original hierarchical triple systems, in which the inner binary has merged and evolved into the sdB star, and the outer (current sdB+MS) binary was never tidally interacting and is thus irrelevant to the production of the sdB. In the \citet{han2002, han2003} merger scenario, two helium white dwarfs merge to make an object capable of core helium burning, and the resulting range of masses for the sdB star is 0.3-0.8$\,\rm{M_{\odot}}$. In the new formation channel proposed by \citet{CW2011}, a helium white dwarf merges with a low-mass hydrogen-burning star whose resultant total mass is $\sim$0.6$\,\rm{M_{\odot}}$. Depending on the mixing history, this object has either a pre-formed helium core or is helium-enriched throughout, and the star experiences a greatly accelerated evolution to the tip of the red giant branch. Only minimal loss is required to remove the residual hydrogen envelope from this object at the time of normal degenerate helium ignition, so the expected mass range for the sdB star in this channel is narrow and at the "canonical" value $\sim$0.5$\,\rm{M_{\odot}}$. There is ample room inside the $\sim$760 day orbit of the present-day \pg\ binary system to accomodate an inner binary which underwent such evolution, and there is no cause for the outer orbit to have been circularised. 
The \citet{CW2011} scenario could account for stars like \pg\ with long (and possibly eccentric) binary orbits. Stable RLOF on the other hand predicts perfectly circular orbits.

\section{Conclusions}

With an orbital period of 760 $\pm$ 4.7 days, \pg\ is the first long period sdB+MS system for which a period has been determined. The spectral type of the companion was found to be a K5 dwarf star, consistent with the mass ratio $M_{\rm MS}/M_{\rm sdB} = 1.6 \pm 0.2$ derived from the radial velocity amplitudes of both stars. However, one has to note that the stated numbers are only indicative of the true orbital parameters, since they are sensitive to the exact uncertainties assigned to the RV data and the assumption of a circular versus eccentric orbit.

At first sight \pg\ is a good candidate for formation through stable Roche lobe overflow if the orbit can be demonstrated to be circular. The predicted number of sdB binaries formed through this channel amounts to the largest contribution of sdB binaries according to binary population synthesis calculations \citep{han2002, han2003}. At the same time, the number of known binaries that have been confirmed as having formed through this channel is small/nil compared to those formed through the common envelope channels. Alternatively, if the common envelope is governed by the gamma-formalism for the common envelope, as used in \citet{nel2010}, there is a population of long period post-common envelope binaries with low-mass secondaries. Thus observing more of these binaries can constrain the first phase of mass transfer. 

If the orbit turns out to be eccentric, the present binary may instead be the remnant of a hierarchical triple-star progenitor system, as outlined by \citet{CW2011}.
Further observations are needed to better establish the orbit of \pg. 

\normalsize{}

\section*{Acknowledgments}

TRM and CMC was supported under a Science and Technology Facilities Council (STFC) rolling grant during the course of this work.
RH\O\ has received funding from the European Research Council under the European Community's Seventh Framework Programme (FP7/2007--2013) /ERC grant agreement N$^{\underline{\mathrm o}}$\,227224 ({\sc prosperity}), as well as from the Research Council of K. U. Leuven grant agreement GOA/2008/04.
LM-R was supported by a PPARC post-doctoral grant and by NWO-VIDI grant 639.042.201 to P. J. Groot during this work. R.A.W. and M.A.S. gratefully acknowledge support from NSF grant AST-0908642. The Isaac Newton and William Herschel Telescopes are operated on the Island of La Palma by the Isaac Newton Group in the Spanish Observatorio del Roque de los Muchachos of the Instituto de Astrof\'isica de Canarias. We thank PATT for their support of this program. We also thank the ING service scheme for obtaining some of the data. This paper uses observations made at the South African Astronomical Observatory (SAAO). A portion of the data reported herein was obtained with the Hobby-Eberly Telescope (HET), which is a joint project of the University of Texas at Austin, the Pennsylvania State University, Stanford University University, Ludwig-Maximillians-Universit\"atM\"unchen, and Georg-August-Universit\"at G\"ottingen. The HET is named in honor of its principal benefactors, William P. Hobby and Robert E. Eberly. This paper also uses observations made with the Nordic Optical Telescope, operated on the island of La Palma jointly by Denmark, Finland, Iceland, Norway, and Sweden, in the Spanish Observatorio del Roque de los Muchachos of the Instituto de Astrof\'isica de Canarias.\\

\normalsize{}
\appendix
\section{RV measurements with error and offset correction}
In table \ref{tbl:fitrvs} we report the radial velocity values for \pg\ with error and offset correction, as used for the final orbital fit results.

\begin{table*}
\centering
\begin{minipage}{160mm}
\caption{Radial velocity measurements for PG\,1018--047 with error and offset correction, as used for the final orbital fit results. }\label{tbl:fitrvs}
\centering
\begin{tabular*}{14cm}{p{2cm}p{2cm}p{2cm}p{2cm}p{2cm}p{2cm}}\hline\hline
HJD			&	RV		&	HJD		&	RV		&	HJD		&	RV		\\
-2450000		&	(km/s)	&	-2450000	&	(km/s)	&	-2450000	&	(km/s)	\\\hline
\multicolumn{2}{c}{INT-B}			& 4197.52858		& 44.4$\pm$  2.9	& \multicolumn{2}{c}{WHT-B}			\\
\multicolumn{2}{c}{\underline{$\sigma$= 2.15 km s$^-1$, off =  + 0.85 km s$^{-1}$}}	& 4547.37872	& 33.9$\pm$  2.7			& \multicolumn{2}{c}{\underline{$\sigma$= 3.40 km s$^-1$, off =  -2.01 km s$^{-1}$}}	\\
2390.51113 	& 35.5 $\pm$ 17.2	& 4547.40665		& 32.0$\pm$  2.7	& 3772.55466		& 33.5$\pm$   3.5	\\
2390.52179 	& 33.6 $\pm$ 9.2	& 4547.62144		& 30.4$\pm$  2.9	& 3772.68565		& 32.5$\pm$   5.4	\\
2391.37270 	& 31.0 $\pm$ 4.8	& 4547.64844		& 33.2$\pm$  3.0	& 4169.43214		& 41.8$\pm$   3.7	\\
2391.38337 	& 33.6 $\pm$ 4.2	& 4902.42320		& 35.7$\pm$  4.9	& 4169.44290		& 45.2$\pm$   3.7	\\
2392.36580 	& 29.2 $\pm$ 3.0	& 4902.44418		& 37.5$\pm$  5.4	& 4169.45704		& 54.5$\pm$   4.0	\\
2392.37993 	& 25.3 $\pm$ 3.0	& \multicolumn{2}{c}{INT-R}			& 4169.46774		& 50.6$\pm$   7.0	\\
2739.51074	& 45.9 $\pm$ 4.2	& \multicolumn{2}{c}{\underline{$\sigma$= 2.15 km s$^-1$, off =  + 0.85 km s$^{-1}$}}	& 4189.53609		& 40.5$\pm$   3.6	\\
2739.52029	& 47.5 $\pm$ 3.9	& 1646.47135		& 32.1$\pm$  4.9	& 4189.547112		& 40.9$\pm$   3.6	\\
2740.44514	& 50.6 $\pm$ 39	& 1646.47650		& 21.3$\pm$  4.8	& 4190.40982		& 44.0$\pm$   3.5	\\
2740.45465	& 52.7 $\pm$ 3.7	& 1654.46074		& 25.6$\pm$  4.3	& 4190.42422		& 48.7$\pm$   3.5	\\
2741.50896	& 50.5 $\pm$ 3.7	& 1654.46780		& 32.9$\pm$  4.3	& 4191.59677		& 42.1$\pm$   3.6	\\
2741.51846	& 50.3 $\pm$ 3.7	& 1977.48220		& 50.4$\pm$  5.7	& 4588.35123		& 27.7$\pm$   4.3	\\
2741.52795	& 50.6 $\pm$ 3.7	& 1978.62533		& 56.0$\pm$  4.4	& 4588.35648		& 27.5$\pm$   4.2	\\
2741.53909	& 54.2 $\pm$ 3.7	& 1979.54044		& 50.8$\pm$  3.9	& 4952.38360		& 53.1$\pm$   4.4	\\
2741.54857	& 49.7 $\pm$ 4.0	& 1979.59958		& 66.3$\pm$  4.0	& 4952.39763		& 47.6$\pm$   4.1	\\
2741.55807	& 49.0 $\pm$ 4.8	& 1979.60655		& 54.8$\pm$  4.0	& 4952.41340		& 56.4$\pm$   3.9	\\
2743.42040	& 51.3 $\pm$ 4.9	& 1982.51489		& 53.0$\pm$  4.4	& 4952.42802		& 56.4$\pm$   3.8	\\
2743.43451	& 61.1 $\pm$ 5.0	& 1982.52185		& 55.5$\pm$  4.6	& \multicolumn{2}{c}{WHT-R}			\\
2743.52128	& 53.0 $\pm$ 6.6	& 1982.55421		& 58.7$\pm$  4.5	& \multicolumn{2}{c}{\underline{$\sigma$= 3.40 km s$^-1$, off =  -2.01 km s$^{-1}$}}	\\
2744.35895	& 55.2 $\pm$ 4.1	& 1982.56118		& 53.4$\pm$  4.4	& 3772.55467		& 24.7$\pm$   3.6	\\
2744.37308	& 50.5 $\pm$ 3.6	& 1982.60059		& 58.4$\pm$  4.4	& 4189.53609		& 47.9$\pm$   3.8	\\
2744.39186	& 50.2 $\pm$ 3.5	& 1982.60755		& 54.0$\pm$  4.5	& 4189.54710		& 50.1$\pm$   3.7	\\
2744.40598	& 52.3 $\pm$ 3.5	& 2031.36769		& 62.9$\pm$  8.6	& 4190.40980		& 49.9$\pm$   3.6	\\
2744.42529	& 54.7 $\pm$ 3.3	& 2031.37466		& 42.2$\pm$  8.7	& 4190.42420		& 51.5$\pm$   3.6	\\
2744.43941	& 53.1 $\pm$ 3.3	& 2031.38641		& 50.3$\pm$  4.9	& 4191.59677		& 50.1$\pm$   3.7	\\
2745.37982	& 49.6 $\pm$ 4.1	& 2032.48243		& 55.7$\pm$  3.7	& 4588.35125		& 32.2$\pm$   4.4	\\
2745.39394	& 46.1 $\pm$ 3.7	& 2032.49633		& 54.5$\pm$  4.1	& 4588.35646		& 20.7$\pm$   4.3	\\
2745.40987	& 52.2 $\pm$ 3.5	& 2033.39672		& 54.7$\pm$  4.6	& 4952.38359		& 52.3$\pm$   6.2	\\
2745.42399	& 52.7 $\pm$ 3.4	& 2033.40369		& 54.6$\pm$  4.8	& 4952.39761		& 51.6$\pm$   5.2	\\
2746.44131	& 55.4 $\pm$ 3.3	& 2037.45584		& 54.9$\pm$  4.4	& 4952.41400		& 49.8$\pm$   4.5	\\
2746.45544	& 51.8 $\pm$ 3.3	& 2037.46282		& 51.8$\pm$  4.1	& 4952.42802		& 49.1$\pm$   4.1	\\
2746.47412	& 56.2 $\pm$ 3.1	& \multicolumn{2}{c}{SAAO}			& \multicolumn{2}{c}{NOT}			\\
2746.48824	& 45.7 $\pm$ 3.2	& \multicolumn{2}{c}{\underline{$\sigma$= 0 km s$^-1$, off =  + 4.63 km s$^{-1}$}}		& \multicolumn{2}{c}{\underline{$\sigma$= 0 km s$^-1$, off =  0 km s$^{-1}$}}	\\
4186.54793	& 38.8 $\pm$ 4.2	& 2364.39275		& 27.7$\pm$  6.0	& 5295.96094		& 35.0$\pm$   1.58	\\
4186.55810	& 40.9 $\pm$ 12.0	& 2364.40338		& 22.5$\pm$  6.6	& \multicolumn{2}{c}{HET}			\\
4188.41925	& 40.8 $\pm$ 3.6	& 3095.38227		& 31.5$\pm$  6.6	& \multicolumn{2}{c}{\underline{$\sigma$= 0 km s$^-1$, off =  0 km s$^{-1}$}}	\\
4188.43328	& 42.9 $\pm$ 3.6	& 3095.40336		& 23.6$\pm$  6.6	& 4441.00811 		& 39.68$\pm$   2.0	\\
4193.37192	& 52.2 $\pm$ 8.1	& 3098.33704		& 17.5$\pm$  9.8	& 4469.93158 		& 36.00$\pm$   2.0	\\
4195.35718	& 42.4 $\pm$ 2.9	& 3101.31890		& 23.4$\pm$  9.2	& 4502.85094 		& 33.50$\pm$   2.0	\\
4195.37121	& 37.2 $\pm$ 2.7	& 3101.32956		& 24.3$\pm$  8.7	& 4550.71405 		& 26.68$\pm$   2.0	\\
4195.61447	& 36.3 $\pm$ 3.3	& 3545.21953		& 56.0$\pm$  7.2	& 4562.68846  		& 29.60$\pm$   2.0	\\
4195.62457	& 39.0 $\pm$ 4.9	& 3546.20756		& 57.2$\pm$  5.8	& 5202.92761  		& 36.47$\pm$   2.0	\\
4197.51456	& 44.6 $\pm$ 2.8	& 				&				& 5278.71693 		& 32.75$\pm$   2.0	\\\hline\hline
\end{tabular*}
\end{minipage}
\end{table*}

\bsp

\label{lastpage}

\end{document}